\documentclass{aa}
\usepackage{graphicx}
\usepackage{epsfig}
\usepackage{latexsym}
\usepackage{natbib}

\usepackage{txfonts}

\def \deg      {$^{\circ}$}

\def \sig      {$\sigma$}
\def \sm       {$\sim$}
\def \gray     {$\gamma$-ray}
\def \grays    {$\gamma$-rays}

\newcommand{\be}{\begin{equation}}
\newcommand{\ee}{\end{equation}}

\begin{document}

   \title{LS 5039 -- the counterpart of the unidentified MeV source GRO~J1823-12}

   \author{W.~Collmar \inst{1},
           S. Zhang \inst{2} 
           }

   \offprints{W. Collmar}
   \mail{wec@mpe.mpg.de}

   \institute{
        Max-Planck-Institut f\"ur extraterrestrische Physik,
        Giessenbachstrasse, D-85748 Garching
      \and
        Key Laboratory for Particle Astrophysics, Institute of High
        Energy Physics, Beijing 100049, China
    }

   \date{accepted: February 6, 2014}

  \abstract
  {
The COMPTEL experiment on CGRO observed the \gray\ sky at energies  
from 0.75 MeV to 30 MeV between April 1991 and June 2000. 
COMPTEL detected many \gray\ sources, among them an unidentified one labeled 
GRO~J1823-12. It is located near l/b$=$17.5\deg/-0.5\deg\ and positionally consistent
with the prominent \gray\ binary LS~5039. 
}
  {
LS~5039 was established as a \gray\ source at TeV energies by HESS and at GeV energies 
by Fermi/LAT during recent years, whose \gray\ radiation is modulated along its 
binary orbit. Given this new information we reanalysed the COMPTEL data of 
GRO~J1823-12 including an orbital resolved analysis.  
}
  {
We applied the standard methods, proper event selections and data binning with 
subsequent maximum-likelihood deconvolution, to analyse the COMPTEL data. In addition 
we developed a tool to select and bin the COMPTEL data in a phase-resolved manner.
We present the orbit-averaged as well as orbit-resolved MeV analyses, 
lightcurves and spectra, and put them into multifrequency context.
}
  {
The COMPTEL data show a significant MeV source, which is positionally consistent with 
LS~5039, however also with other closeby Fermi/LAT sources. The orbit-resolved analysis
provides strong evidence, at about the 3\sig\ level, that the MeV flux of GRO~J1823-12 
is modulated along the binary orbit of about 3.9 days of LS~5039. We show that at MeV 
energies the source is brighter at the orbital part around the inferior conjuction than 
at the part of the superior conjunction, being in phase with X-rays and TeV \grays,  
however being in anti-phase with GeV \grays. The high-energy SED (X-rays to 
TeV \grays) shows the high-energy emission maximum of LS~5039 at MeV energies.    
  }
  {
We conclude that the COMPTEL source GRO~J1823-12 is the counterpart of the 
microquasar LS~5039, at least for the majority of its MeV emission. The COMPTEL 
fluxes, put into multifrequency perspective, provide new constraints on the 
modelling of the high-energy emission pattern of the \gray\ binary LS~5039.
}

   \keywords{gamma-rays: observations --
             binaries: individual: LS~5039/GRO~J1823-12
            }

   \authorrunning{W. Collmar et al.}
   \titlerunning{LS 5039 -- the counterpart of the unidentified MeV source GRO~J1823-12} 
   \maketitle
%

\section{Introduction}
The Compton Gamma-Ray Observatory (CGRO) observed the universe in \gray\ energies 
with unprecedented sensitivity for more than 9 years between its launch in 
April 1991 and its reentry into the earth atmosphere in June 2000. 
CGRO carried the four \gray\ experiments BATSE, OSSE, COMPTEL and EGRET,
which provided many new and exciting results in the regime of the \gray\ band, accesible only 
from space. In particular, many point sources between \sm 1~MeV and \sm10~GeV were
detected, of which surprisingly a big fraction remained unidentified until the 
end of the CGRO mission. For example, out of the 271 EGRET \gray\ sources listed in the 
3rd EGRET source catalogue for energies above 100~MeV, 171 remained unidentified 
\citep{Hartman99}. 

The COMPTEL experiment was sensitive to soft \grays\ (0.75-30~MeV) and finally 
opened the MeV-band, basically unexplored before CGRO, as a new astronomical window.
Apart from \gray\ bursts and AGN, the majority of the COMPTEL sources
are unidentified objects. The first COMPTEL source catalogue lists 10 AGN and 9 unidentified \gray\ sources
\citep{Schoenfelder00}. One of these unidentified sources is GRO~J1823-12, 
which is significantly visible in time-averaged COMPTEL maps of all 
standard energy bands between 1 and 30 MeV \citep[e.g.][] {Collmar00a}.  
GRO~J1823-12 is located near l/b$=$17.5\deg/-0.5\deg, spatially consistent with 
three EGRET \gray\ sources, the most prominent one being 3EG~J1824-151 which was proposed as the 
counterpart of the microquasar LS~5039 \citep{Paredes00}. Results of  preliminary analyses 
of the COMPTEL data on GRO~J1823-12 between 1991 and 1997 were reported by 
\citet{Strong01} and \citet{Collmar03}. 
The source showed a hard power-law spectrum with photon index of \sm1.6 in the energy band 
between 1-30~MeV. However, due to the large error location area of the COMPTEL source, 
the source remained unidentified.      

LS~5039 \citep{Stephenson71}, a luminous star of the southern 
milky way (l/b$=$16.88\deg/-1.29\deg),
was first identified as a high-energy source by a cross-correlation with 
unidentified ROSAT X-ray sources \citep{Motch97}. They suggested
LS 5039 to be a high-mass X-ray binary system. 
The detection of LS~5039 by the High Energetic Spectroscopic System (HESS) 
at energies above 250 GeV \citep{Aharonian05} proved LS~5039 being even a \gray\ source. 
The HESS measurements showed that the flux and the energy spectrum
of LS~5039 is modulated with the orbital period of the binary system of \sm3.9 days
\citep{Aharonian06a}.  
At present LS~5039 is well established as being a 
high-energy source, e.g. at energies above 100~MeV by Fermi/LAT \citep{Abdo09, Hadasch12}, 
at hard X-rays (25 - 200~keV) by INTEGRAL \citep{Hoffmann09} and at X-ray energies 
(1 - 10~keV) by e.g. {\it Suzaku} \citep{Takahashi09}. 
The microquasar is significantly detected at these energies, including 
the orbital modulation of its flux and energy spectrum. 

Because this peculiar time variability of LS~5039 became known during last years
and was detected in COMPTEL's neighboring energy bands, we reanalysed our COMPTEL 
data on GRO~J1823-12. The detection of such a time signature in the COMPTEL MeV data would 
(at least partly) identify this COMPTEL source as counterpart of LS~5039. 
In this paper we report the analysis results and discuss them.

The paper is organized as follows: in Sect. 2 we briefly describe the COMPTEL 
instrument and the applied data analysis methods, the analysed observations are summarized 
in Sect. 3. In Sect. 4 we present the analysis results and a high-energy SED in Sect.~5. 
In Sect. 6 we discuss our results and finally summarise and conclude in Sect.~7.


\section{\label{instrument}Instrument and data analysis}

The imaging Compton Telescope COMPTEL was sensitive to $\gamma$-rays in 
the energy range 0.75-30 MeV with an energy-dependent energy
and angular resolution of 5$\%$ - 8$\%$ (FWHM) and 
1.7\deg\ - 4.4\deg\ (FWHM), respectively. 
It had a large field of view of about 1 steradian and was able
to detect $\gamma$-ray sources with a location accuracy of the order
of 1$^{\circ}$ - 2$^{\circ}$, depending on source flux. For the 
details about COMPTEL see \citet{Schoenfelder93}.

COMPTEL contained two detector arrays in which an incident $\gamma$-ray
photon is first Compton scattered in a detector of the upper detector array
and -- in the favorable case -- then interacts with a detector of the
lower detector array. The scattered photon direction ($\chi$,$\psi$)
is obtained from the interaction locations in the two detectors. 
The Compton scatter angle $\bar{\varphi}$ is calculated from the
measured energy deposits of the photon in the two detectors. 
These quantities, scatter direction and angle, constitute a
three-dimensional data space in which the spatial response of
the instrument is cone-shaped and standard imaging methods, e.g. 
maximum entropy and maximum likelihood, are applied.
In the COMPTEL data analysis package the maximum-likelihood method
is used to estimate source parameters like detection 
significances, fluxes and flux errors. 
The detection significance is calculated from the 
quantity -2ln$\lambda$, where $\lambda$ is the ratio of the
likelihood L$_{0}$ (background) and the likelihood L$_{1}$
(source + background). The quantity -2ln$\lambda$ has
a $\chi_{3}^{2}$ distribution (3 degrees of freedom) 
for a unknown source and a $\chi_{1}^{2}$ distribution for 
a known source \citep{Boer92}.
The instrumental COMPTEL background was modelled by the 
standard filter technique in data space \citep{Bloemen94}.

To account for any diffuse astrophysical background in the data, 
the COMPTEL software provides three diffuse all-sky emission models,
which can be fit additionally to the data by a global scaling factor.
An uniform isotrop model represents the extra-galactic \gray\ 
background. Two models are available to account for the emission of the galaxy: 
One model represents the galactic inverse-Compton emission while the 
other the galactic bremsstrahlung emission via galactic density profiles of HI and CO.       

Due to the location of GRO~J1823-12 at low galactic-latitudes 
in the inner galaxy (l/b=\sm17.5\deg/-0.5\deg) we included all three diffuse 
emission models in the source analysis. The scaling factor of the 
extragalactic \gray\ background was fixed to the fluxes calculated from the 
diffuse extra-galactic MeV-spectrum as derived by \citet{Weidenspointner99}. 
To account for the specific environment of the analysed sky regions, the scaling 
factors for the two galactic diffuse emission models were allowed to vary freely. However, 
we checked these scaling factors numerically after the fitting procedure in order to 
avoid unreasonable/unphysical values which may lead to unrealistic source fluxes. 
In addition we added known nearby COMPTEL point sources in the fitting process. So the 
scaling parameters of the diffuse galactic emission models as well as the 
point-source fluxes were estimated by simultaneous fitting. 

Because in sky maps the location of the microquasar LS~5039 was always close 
to the maximum of the likelihood-distribution, we assumed for the flux estimates 
a point source at the LS~5039 location. The most important additional source is the 
``nearby'' quasar PKS~1830-210 at l/b=\sm12.2\deg/-5.7\deg,
 which is also a known COMPTEL source \citep[e.g.][]{Zhang08}. The inclusion of 
two further MeV source, one at the Galactic Center,
the other one being the COMPTEL detected blazar PKS~1622-297 \citep{Zhang02} showed 
no influence on the derived source fluxes of GRO~J1823-12. 

\begin{figure*}[t]
\centering
   \includegraphics[width=0.33\textwidth]{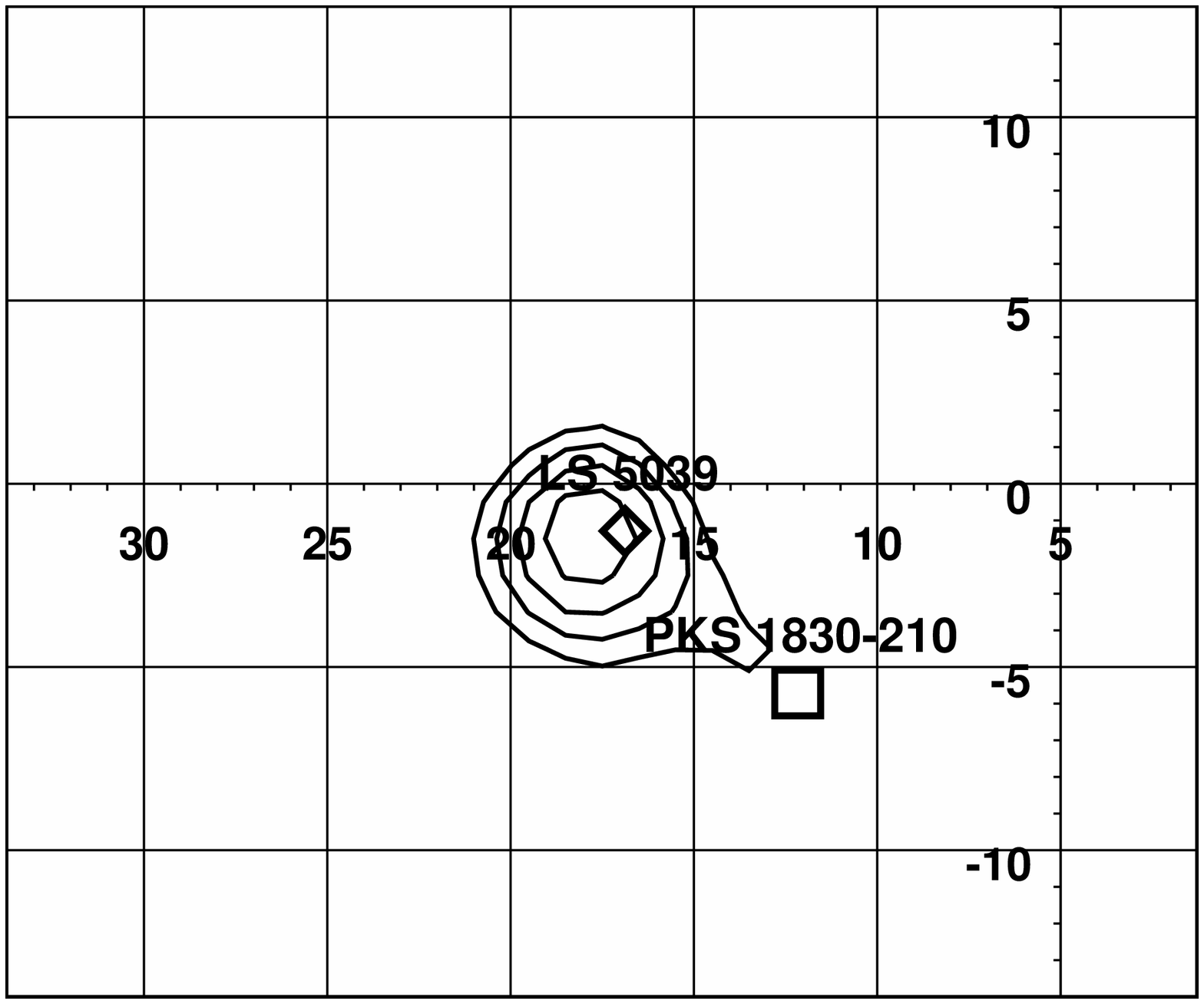} \hfill
   \includegraphics[width=0.33\textwidth]{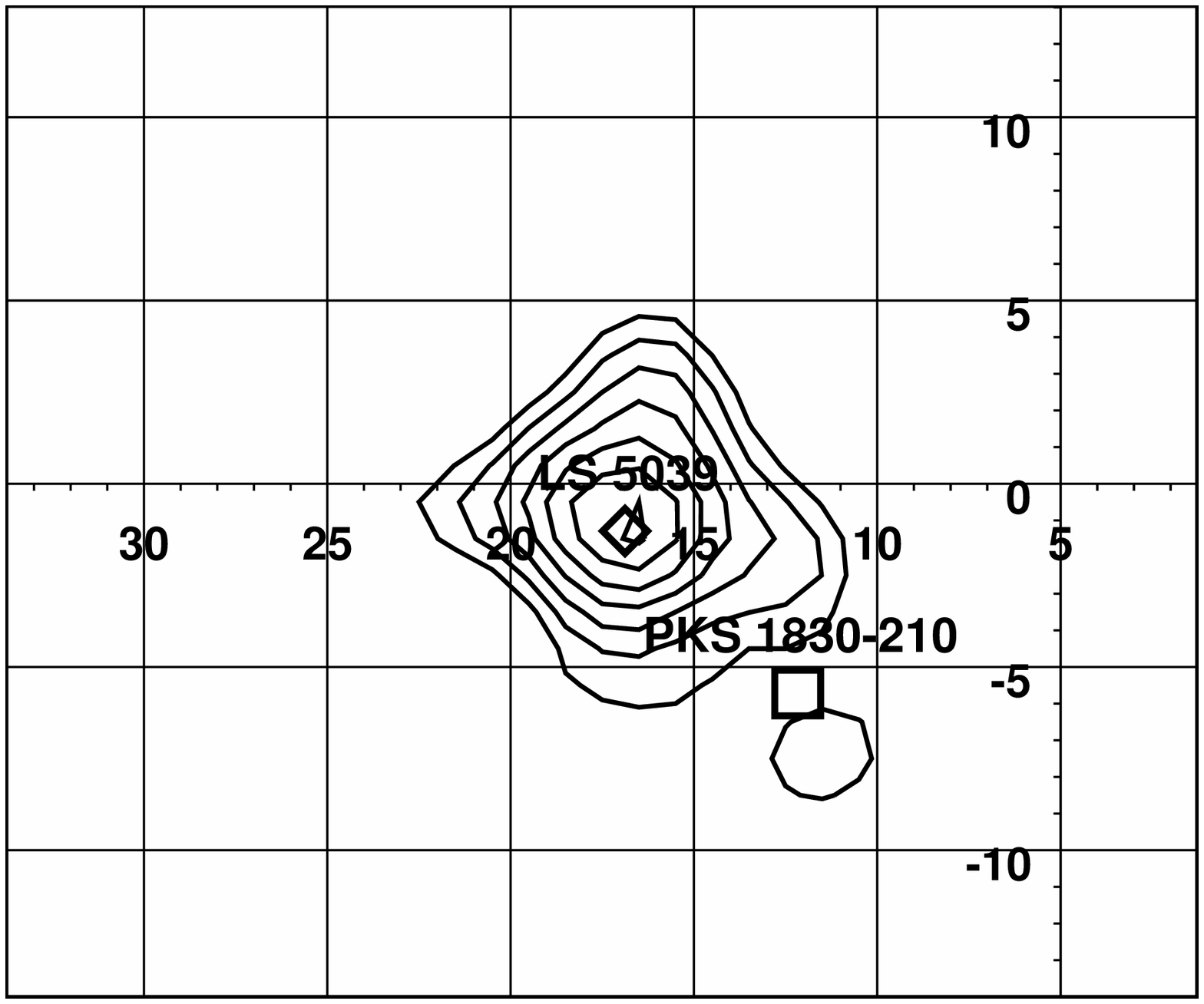} \hfill
   \includegraphics[width=0.33\textwidth]{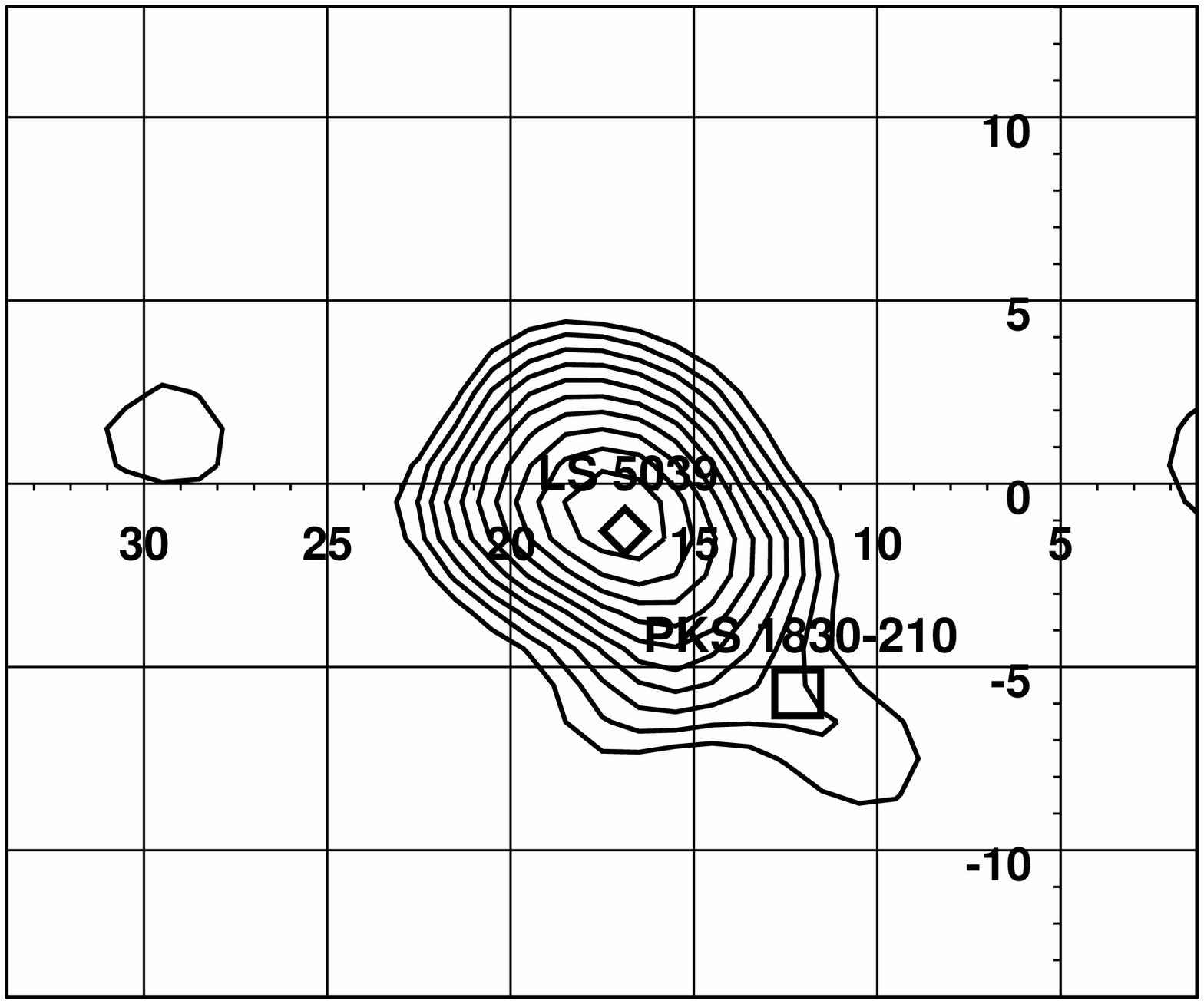}
\caption{
COMPTEL 1-3 (left), 3-10 (middle) and 10-30 MeV (right) maps, generated in galactic coordinates (l, b), 
show the statistical evidence for GRO~J1823-12 for the sum of all data. The contour lines start at a detection
significance of 3\sig\ (1 d.o.f. for a known source) with a step of 0.5\sig.
The location of the microquasar LS~5039 (diamond, l/b: 16.88\deg/-1.29\deg) and 
the lensed quasar PKS~1830-210 (square) 
are indicated, i.e. the locations of the simultaneously fitted COMPTEL sources. 
The quasar PKS~1622-297 and a possible source at the Galactic Center, both out of the plot region, 
were also taken into account in the simultaneous fitting. The 10-30 MeV band yields the highest 
detection significance (7.9\sig) at the position of LS~5039.   
}
\label{ls5039_map}
\end{figure*}

We analysed the data in 3 of the 4 so-called ``standard'' COMPTEL energy bands 
(0.75-1, 1-3, 3-10, and 10-30~MeV). We omitted the lowest COMPTEL band (0.75-1~MeV) 
because the systematics in this narrow COMPTEL band 
(small efective area) at the lower edge of the COMPTEL energy range are large 
and not well understood. 

To estimate the flux of sources in the 3 bands, we applied 
instrumental point spread functions assuming an E$^{-2}$ power-law shape 
for the source input spectrum in the different
energy bands. We note that the derived fluxes are weakly
dependent on this particular shape, e.g. the flux 
changes by applying e.g. a E$^{-1.6}$ power-law  
shape PSFs were always significantly smaller than the 1\sig\ errors on the estimated source 
fluxes by application of an E$^{-2}$ power-law PSF. 

The goal of the analyses was to derive the MeV-properties of the unidentfied COMPTEL source 
near l/b=\sm17.5\deg/-0.5\deg. Therefore we selected all useful COMPTEL observations within 
a pointing direction of 35\deg\ to this sky position (see Sect.~3). We analysed these data
in time periods of at least 1 year. The systematic uncertainties on the diffuse emission models 
make flux estimates for individual pointings to this sky region, i.e. 1 to 2 weeks, unreliable.
In fact, we present quantitative spectral results only for longer time periods, in particular 
for the sum of all data, i.e. summing up the complete mission, thereby using the maximal available 
data statistics.

We like to note that we checked our analysis results by analysing the data also for
different data selections. By analysing selected sub-periods in time as well as analysing 
the data in revised energy bands, we yield consistent analysis results in flux levels and source 
significances. This shows that the results presented in this paper are independent of the applied
analysis mode or the selected observational period.

\section{\label{observations}Observations}

During the complete CGRO mission from April 1991 to June 2000,
GRO~J1823-12 was in 51 observational periods, so-called CGRO viewing periods (VPs),
within 35\deg\ of the COMPTEL pointing direction. Each VP lasts typically for 1 to 2 weeks.   
Table~\ref{obs} provides the basic informations on these 51 VPs and groups them 
within the 9 longterm observational periods of CGRO, so-called Phases or Cycles. 
A CGRO Phase or Cycle covers a time period of roughly 1 year. 
The CGRO mission was subdivided into four mission Phases of which the last one consists 
of six (4 to 9) Cycles. COMPTEL observations
on the GRO J1823-12 region are available up to last Cycle, i.e. up to early 2000. 
These 51 VPs add up to a total effective exposures  -- COMPTEL 
pointing directly and uninterruptedly to the source -- of \sm 81 days.
We have analysed the sum of these 51 VPs and present the results in Sect.~\ref{results} 
of this paper.   

\begin{table*}[hb]
\caption{COMPTEL observational periods of the GRO J1823-12 region during the  
CGRO mission, where the MeV source was within 35\deg\ of
the pointing direction. The CGRO viewing periods (VPs), their time periods,
prime observational targets, pointing offset angles,
and the CGRO Phases including their effective exposures (days) are given.
The values are calculated for the sky location l/b:17.0\deg/-1.0\deg.
}
\begin{flushleft}
\begin{tabular}{cccccccc}
\hline 
\multicolumn{1}{c}{VP}&\multicolumn{1}{c}{Date}&\multicolumn{1}{c}{Target}&\multicolumn{1}{c}{Ang.Sep.}&\multicolumn{1}{c}{Duration}&\multicolumn{1}{c}{CGRO Phase}\\  
\multicolumn{1}{c}{$\#$}&\multicolumn{1}{c}{ }&\multicolumn{1}{c}{ }&\multicolumn{1}{c}{degs}&\multicolumn{1}{c}{days}&\multicolumn{1}{c}{}\\ \hline
5.0   & 12/07/91-26/07/91 & Gal. Center   & 17.2$^{\circ}$&  14 & Phase I\\ 
7.5   & 15/08/91-22/08/91 & Gal. 25-14    & 15.2$^{\circ}$&  7 &\\ 
13.0  & 31/10/91-07/11/91 & Gal. 25-14    & 15.2$^{\circ}$&  7 &\\ 
16.0  & 12/12/91-27/12/91 & Sco X-1       & 26.8$^{\circ}$& 15 &\\ 
20.0  & 06/02/92-27/12/91 & SS 433        & 23.1$^{\circ}$& 14 &\\ 
43.0  & 29/10/92-03/11/92 & Mrk 509       & 30.1$^{\circ}$&  5 & 13.06\\ \hline 
210.0 & 22/02/93-25/02/93 & Gal. Center   & 22.1$^{\circ}$&  3 & Phase II\\ 
214.0 & 29/03/93-01/04/93 & Gal. Center   & 22.1$^{\circ}$&  3 &\\
219.4 & 05/05/93-06/05/93 & Gal. Center   & 31.6$^{\circ}$&  1 &\\
223.0 & 31/05/93-03/06/93 & 1E 1740-29    & 18.0$^{\circ}$&  3 &\\
226.0 & 19/06/93-29/06/93 & Gal. 355+5    & 22.8$^{\circ}$& 10 &\\
231.0 & 03/08/93-10/08/93 & NGC 6814      & 13.1$^{\circ}$&  7 &\\
229.0 & 10/08/93-11/08/93 & Gal. 5+5      & 13.4$^{\circ}$&  1 &\\
229.5 & 12/08/93-17/08/93 & Gal. 5+5      & 13.4$^{\circ}$&  5 &\\
232.0 & 24/08/93-26/08/93 & Gal. 348+0    & 29.0$^{\circ}$&  2 &\\
232.5 & 26/08/93-07/09/93 & Gal. 348+0    & 29.0$^{\circ}$& 12 & 5.91 \\ \hline
302.3 & 09/09/93-21/09/93 & GX 1+4        & 18.8$^{\circ}$& 12 & Phase III \\
323.0 & 22/03/94-05/04/94 & Gal. 357-11   & 22.2$^{\circ}$& 14 &\\
324.0 & 19/04/94-26/04/94 & Gal. 016+05   &  7.3$^{\circ}$&  7 &\\
330.0 & 10/06/94-14/06/94 & Gal. 018+00   &  1.4$^{\circ}$&  4 &\\
332.0 & 18/06/94-05/07/94 & Gal. 018+00   &  1.4$^{\circ}$& 17 &\\
334.0 & 18/07/94-25/07/94 & Gal. 009-08   & 10.6$^{\circ}$&  7 &\\
338.0 & 29/08/94-31/08/94 & GRO J1655-40  & 32.1$^{\circ}$& 14 & 15.29 \\ \hline
414.3 & 29/03/95-04/04/95 & GRO J1655-40  & 30.1$^{\circ}$&  6 & Phase IV/\\
421.0 & 06/06/95-13/06/95 & Gal. Center   & 22.0$^{\circ}$&  7 & Cycle 4\\
422.0 & 13/06/95-20/06/95 & Gal. Center   & 22.0$^{\circ}$&  7 &\\
423.0 & 20/06/95-30/06/95 & Gal. Center   & 14.0$^{\circ}$& 10 &\\
423.5 & 30/06/95-10/07/95 & PKS 1622-297  & 33.8$^{\circ}$& 10 &\\
429.0 & 20/09/95-27/09/95 & Gal. 018+4    &  5.1$^{\circ}$&  7 & 8.55 \\ \hline
501.0 & 03/10/95-17/10/95 & Gal. Center   & 12.1$^{\circ}$& 14 & Phase IV/\\
508.0 & 14/12/95-20/12/95 & Gal. 005+0    & 10.0$^{\circ}$&  6 & Cycle 5\\
509.0 & 20/12/95-02/01/96 & Gal. 021+14   & 14.9$^{\circ}$& 13 &\\
524.0 & 09/07/96-23/07/96 & GX 339-4      & 34.1$^{\circ}$& 14 &\\
529.5 & 27/08/96-06/09/96 & GRO 1655-40   & 32.1$^{\circ}$& 10 & 9.77 \\ \hline
624.1 & 04/02/97-11/02/97 & Gal. 016+00   &  4.1$^{\circ}$&  7 & Phase IV/ \\
619.2 & 14/05/97-20/05/97 & GRS 1915+105  & 30.0$^{\circ}$&  6 & Cycle 6\\
620.0 & 10/06/97-17/06/97 & Gal. 016+4    &  5.1$^{\circ}$&  7 &\\
625.0 & 05/08/97-19/08/97 & GRS 1758-258  & 15.1$^{\circ}$& 14 &\\
615.1 & 19/08/97-26/08/97 & PKS 1622-297  & 31.1$^{\circ}$&  7 & 8.48 \\ \hline
703.0 & 25/11/97-02/12/97 & Gal. 035+20   & 26.7$^{\circ}$&  7 & Phase IV/ \\
704.0 & 02/12/97-09/12/97 & Gal. 035+20   & 23.3$^{\circ}$&  7 & Cycle 7\\
712.0 & 27/01/98-24/02/98 & Gal. 035+20   & 26.2$^{\circ}$& 28 &\\
720.5 & 05/05/98-15/05/98 & GRS 1915+105  & 28.0$^{\circ}$& 10 &\\
737.0 & 24/11/98-01/12/98 & Gal. 044-09   & 28.0$^{\circ}$&  7 & 9.12 \\ \hline
811.5 & 06/04/99-13/04/99 & GRS 1915+105  & 25.5$^{\circ}$&  7 & Phase IV/ \\
812.5 & 13/04/99-20/04/99 & GRS 1915+105  & 25.5$^{\circ}$&  7 & Cycle 8\\
813.5 & 20/04/99-27/04/99 & GRS 1915+105  & 25.3$^{\circ}$&  7 & 3.10 \\ \hline
901.0 & 09/12/99-14/12/99 & Gal. 004+00   & 15.0$^{\circ}$&  5 & Phase IV/ \\
902.0 & 14/12/99-21/12/99 & Gal. 005-05   & 12.6$^{\circ}$&  7 & Cycle 9\\
902.0 & 14/12/99-21/12/99 & Gal. 005-05   & 12.6$^{\circ}$&  7 & \\
904.2 & 28/12/99-04/91/00 & Sun           & 29.1$^{\circ}$&  7 & \\
905.0 & 04/01/00-11/01/00 & Gal. 018-19   & 18.0$^{\circ}$&  7 & \\
906.0 & 11/01/00-19/01/00 & Sun           & 25.2$^{\circ}$&  7 & \\
907.0 & 19/01/00-25/01/00 & Sun           & 30.9$^{\circ}$&  6 & 7.84 \\ \hline
All   & Sum of Mission    &               &               &    & 81.11 \\ \hline

\hline
\end{tabular}\end{flushleft}
\label{obs}
\end{table*}

\section{\label{results}Results}

\subsection{\label{detections}Orbit-averaged analyses}

\subsubsection{Source detections}
In a first step we generated circular skymaps around the estimated sky position of 
GRO~J1823-12 (l/b:17.5\deg/-0.5\deg) by applying the described analysis methods 
and data selections (see Sect. 2 and 3).
These maps with radius of 40\deg\ and a sky binning of 1\deg\ were cross-checked for 
MeV sources. The central part of the three maps for the sum of all 51 VPs, 
i.e. complete mission with an effective COMPTEL 
exposure of $\sim$81~days (see Tab.~\ref{obs}), are shown in Fig.~\ref{ls5039_map}. 
Each map, derived for the COMPTEL standard energy bands 1-3~MeV, 3-10~MeV and 
10-30~MeV, shows a significant source, being consistent with the location of the 
microquasar LS~5039. The most significant sky pixel either contains the location of the 
microquasar (3-10~MeV, -2ln$\lambda$: 24.3) or is 1 pixel apart (1-3~MeV, -2ln$\lambda$: 37.0; 
10-30~MeV, -2ln$\lambda$: 63.9).

\begin{figure}[ht]
\centering
   \includegraphics[width=\columnwidth]{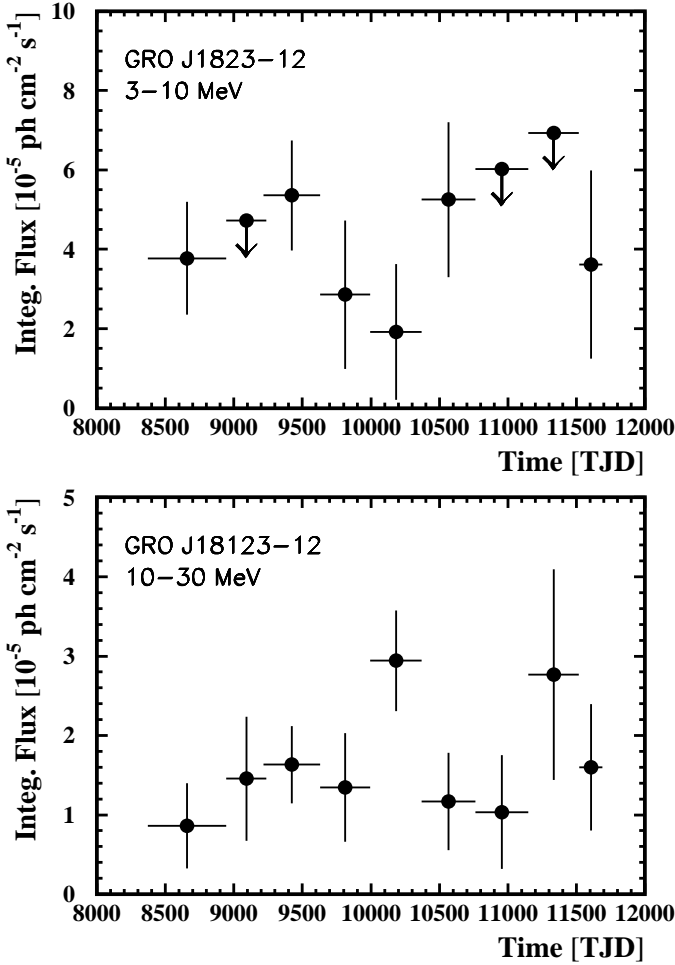}
\caption{
COMPTEL light curves of GRO~J1823-12, fitted at the sky location of 
LS~5039, in the 3-10 and 10-30 MeV bands. The flux points (integral photon fluxes for 
the given bandwidth)  
are averaged over individual CGRO phases/cycles, each covering roughly 1 year. 
For the details on the observations see Tab.~\ref{obs}.
}
\label{ls5039_lco}
\end{figure}

\subsubsection{Time variability}
We investigated the time variability of the MeV-flux on timescales of CGRO Phases, i.e. 
on time periods of the order of about 1 year. Fig.~\ref{ls5039_lco} shows the fluxes 
along the COMPTEL mission in the two uppermost (3-10 and 10-30 MeV) COMPTEL bands for 
the sum of VPs of individual CGRO Phases (see Tab.~\ref{obs}).
We show these two bands, because the background changes 
(e.g. depending on orbital altitude) along the mission are small, 
in particular the 10-30~MeV band is unaffected, which makes these high-energy 
bands more reliable than the lower-energy ones. Also, in these bands GRO~J1823-12 
is more significantly detected in time-averaged analyses. We plot a flux point, if the source
reaches at least a 1\sig-detection level. If not, we plot a 2\sig-upper limit.

The 3-10~MeV lightcurve is less conlusive. Detections, although on a low significance level, 
and non-detections occur along the mission, adding up to a \sm5.5\sig-detection for the sum of 
all data. 
The 10-30 MeV light curve however, shows always evidence (at least 1\sig) for the source with some 
hints for time variability. During CGRO Phase IV/Cycle 5, the flux is more than three times higher 
than during the observations in CGRO Phase I. However, a quantitative analysis
by assuming a constant flux results in a low and insignificant probability of 0.60
for a time variable flux. At these highest COMPTEL energies, 
GRO~J1823-12 seems to be a steady MeV-emitter.
This is in agreement with the observations at GeV- and 
TeV-energies, where the longterm light curves over years are also consistent 
with a steady source \citep[e.g.][]{Hadasch12,Aharonian06a}.

\subsubsection{Energy spectra}

\begin{figure}[th]
\centering
   \includegraphics[width=\columnwidth]{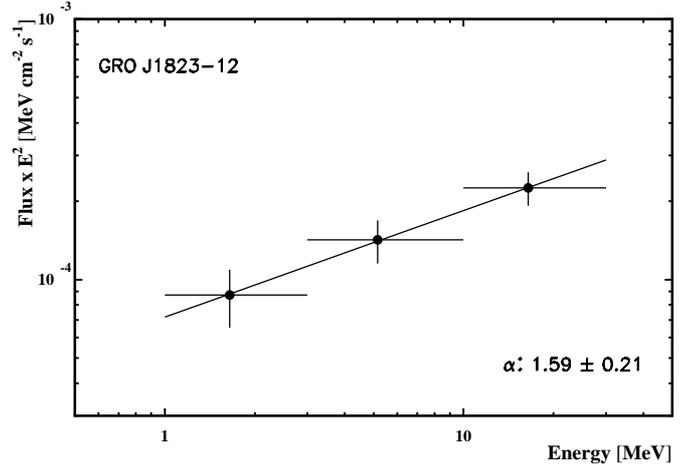}
\caption{
COMPTEL energy  spectrum of GRO~J1823-12, fitted at the location of LS~5039,
in an E$^{2} \times$ differential flux representation  
for the sum of all data. The fluxes are derived in 
the 3 standard COMPTEL energy bands (1-3, 3-10, 10-30 MeV). The error bars are 1\sig. 
The solid line represents the best-fitting power-law shape between 1 and 30 MeV.
}
\label{ls5039_specs}
\end{figure}

To derive the COMPTEL fluxes of GRO~J1823-12, we have applied the standard maximum-likelihood method 
as described in Sect. 2. Background subtracted and deconvolved source fluxes in three standard energy 
bands have been derived by simultaneous fitting of four \gray\ sources, two diffuse models, representing 
the galactic emission, and an isotropic component (flux fixed) describing the extragalactic diffuse emission. 
Since there is no obvious time variability we concentrate on the most significant time-averaged data, 
the sum of all data (i.e. all VPs listed in Table~\ref{obs}). 
Table~\ref{tabflux} gives the corresponding MeV fluxes of GRO~J1823-12, 
assumed to be the microquasar LS~5039, in three bands. 
The corresponding spectrum including the best-fit power-law shape is shown 
in Fig.~\ref{ls5039_specs}. In an E$^{2}$ $\times$ differential flux respresentation, the fluxes 
rise towards higher energies. We fit a simple power-law model,

\begin{equation}
I(E) = I_{0}  (E/E_{0})^{-\alpha} \,\, {\rm photons\,\,cm}^{-2} {\rm s}^{-1} {\rm MeV} ^{-1}
\end{equation}  

where the parameter $\alpha$ is the photon index, and I$_0$ the 
differential flux at the normalization energy E$_0$, which was set to 5~MeV 
throughout all our analyses. We find a well-fitting hard power-law shape
with a photon index of the order of \sm1.6. (Table~\ref{tabspec}).

\begin{table}[htb]
\caption{Fluxes, assuming a source at the location of LS~5039 for the sum of all data 
(VPs 5.0 -- 907.0). 
The flux units are 10$^{-5}$ ph cm$^{-2}$ s$^{-1}$. The energy bands are given in MeV.
The errors bars are 1\sig.
} 
\begin{flushleft}
\begin{tabular}{ccccc}
\hline\noalign{\smallskip}
  Period & 1-3 & 3-10 & 10-30   \\
\hline\noalign{\smallskip}
VPs 5.0 - 907.0 & 5.65$\pm$1.43 & 3.22$\pm$0.60 & 1.46$\pm$0.21 \\
\hline\noalign{\smallskip}
\end{tabular}\end{flushleft}
\label{tabflux}
\end{table}

\begin{table}[htb]
\caption[]{
The result of the power-law fitting of the COMPTEL 
 spectrum between 1 and 30~MeV for the sum of all data (Fig.~\ref{ls5039_specs}). 
The errors are 1\sig\ ($\chi^{2}_{min}$ + 2.3 for 2 parameters of interest).
}
\begin{flushleft}
\begin{tabular}{cccc}
\hline\noalign{\smallskip}
Energy & Photon Index & I$_0$ & $\chi^{2}_{min}$   \\
 & ($\alpha$) & (10$^{-6}$ cm$^{-2}$ s$^{-1}$ MeV$^{-1}$) &   \\ 
\hline\noalign{\smallskip}
1-30 MeV & 1.59$\pm$0.21  & 5.55$\pm$1.05 & 0.01 \\  
\hline\noalign{\smallskip}
\end{tabular}\end{flushleft}
\label{tabspec}
\end{table}

\subsubsection{Fermi sources in the GRO~J1823-12 field}

\begin{figure}[ht]
\centering
   \includegraphics[width=\columnwidth]{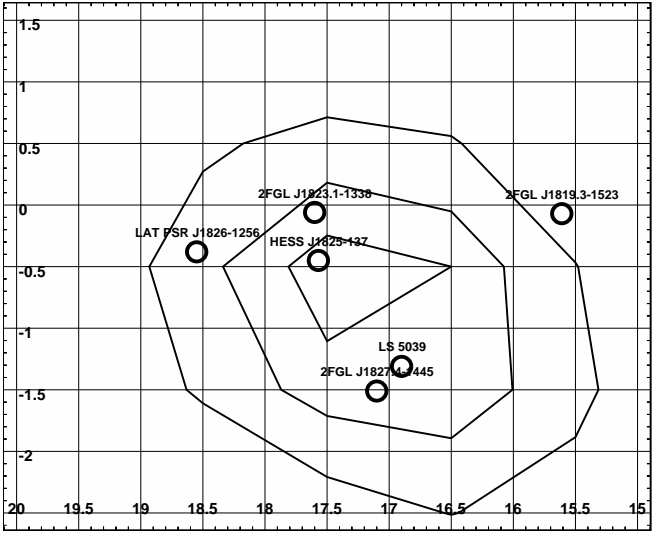}
\caption{
COMPTEL error location contours for the most significant COMPTEL detection of GRO~J1823-12,
obtained in the 10-30 MeV band for the sum of all data. 
The contour lines are plotted on a map in galactic coordinates (l, b) of the LS~5039-region.
The error contours start with 1\sig\ with steps of 1\sig.  The sky positions (circles) of 
all \gray\ source are shown, which are listed in the second Fermi catalog and which are 
within a search radius of 2\deg\ around the pixel center of the best-fit source 
location (l/b:17.5\deg/-0.5\deg).
}
\label{ls5039_errloc}
\end{figure}

Fig.~\ref{ls5039_errloc} shows the error location 1, 2, and 3\sig\ contours around the most 
significant COMPTEL detection, the 10-30~MeV band for the sum of all data. There are 5 
Fermi-detected \gray\ sources located within the 3\sig-confidence contour. LS~5039 is located 
in the sky pixel next to maximum. The source identification is not obvious and so we need 
other means to decide on the counterpart of the COMPTEL source.   

Closest to the significance maximum is the source HESS~J1825-137 \citep{Aharonian06b},
a pulsar wind nebula which is also detected by Fermi/LAT at energies between 1 and 100~GeV
\citep{Grondin11}. Below 1 GeV the source is not detected by Fermi/LAT, consistent with 
the hard spectral power-law index of $\sim$1.38 measured above 1~GeV. The SED, provided 
by \citet{Grondin11}, suggests 
an emission minimum in the COMPTEL band, which makes HESS~J1825-137 an unlikely counterpart 
of the COMPTEL source.

\begin{figure}[ht]
\centering
   \includegraphics[width=\columnwidth]{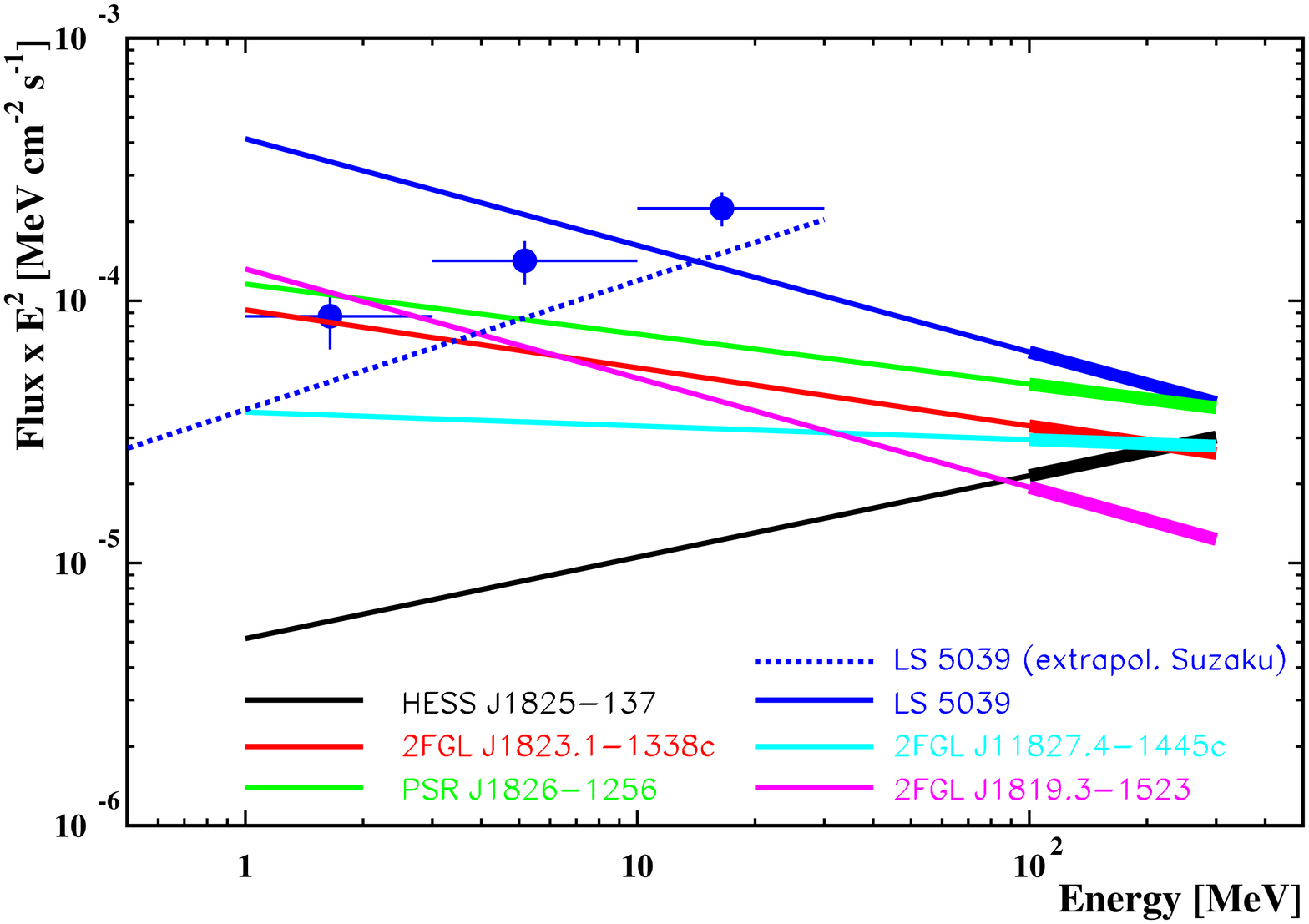}
\caption{
The time-averaged COMPTEL energy spectrum of GRO~J1823-12 for the 
sum of all data (see Fig.~\ref{ls5039_specs}) is
compared to the extrapolations of the Fermi/LAT spectrum of the 6 closest \gray\ sources, 
and to the extrapolation of the time-averaged X-ray spectrum (1-10~keV, dashed blue line) 
of LS~5039 as measured by {\it Suzaku} (Takahasi et al. 2009). 
The thick solid lines represent the spectral shape derived from best-fit Fermi/LAT  
integral fluxes between 100 and 300~MeV by assuming a power-law shape of fixed index. 
This index was derived by a power-law throughout the Fermi/LAT energy band. 
The thin solid lines represent the spectral extrapolations down to 1~MeV. For the 
known Fermi/LAT sources, 
the spectral extrapolation for LS~5039 is coming closest to the COMPTEL 10-30~MeV measurement.
Also the power-law extrapolation of the {\it Suzaku}-measured X-ray spectrum is 
in reasonable agreement with the COMPTEL spectrum of GRO~J1823-12.  
For more details see text.
}
\label{ls5039_CF_specs}
\end{figure}

Apart from LS~5039 none of the other sources is discussed in spectral detail in the literature. 
In order to derive some estimates on their fluxes we took the relevant values from the Fermi/LAT
2nd Source Catalog \citep{Nolan12}. In particular we used the integral photon fluxes from the
100 to 300 MeV band and a fixed power-law shape of the energy spectrum, whose power-law index 
was derived by a likelihood analysis in the band from 100 MeV to 100 GeV \citep{Nolan12}.
The spectral extrapolation for LS~5039 is coming closest to 
the measured COMPTEL flux of GRO~J1823-12 (see Table~\ref{fermi_srcs} and 
Fig.~\ref{ls5039_CF_specs}), reaching a flux value of
0.91 $\times$ 10$^{-5}$ ph cm$^{-2}$ s$^{-1}$ compared to the measured value of 
1.46 $\times$ 10$^{-5}$ ph cm$^{-2}$ s$^{-1}$. In fact the sum of the extrapolated fluxes
of the 6 sources is 2.28 in these units. It should be considered as an upper limit, because 
the Fermi/LAT spectra of some sources have curved shapes, 
``LogParabola'' or ``Power-law with exponential cutoffs''. Assuming a power-law shape 
throughout the band results at the low end to an overestimaton of the flux values. 
Nevertheless, the extrapolated and measured flux values agree withing 
roughly a factor of two. 
The values are given in Table~\ref{fermi_srcs} and their spectral comparison 
to the COMPTEL fluxes is graphically shown in Fig.~\ref{ls5039_CF_specs}. 

\citet{Takahashi09} measured the X-ray spectrum of LS~5039 with {\it Suzaku} between 0.6 and 10~keV 
very accurately. The power-law extrapolation of their time-averaged X-ray spectrum up to the COMPTEL 
energies yields a flux of 1.0$\pm$0.1 $\times$ 10$^{-5}$ ph cm$^{-2}$ s$^{-1}$ 
for the COMPTEL 10-30~MeV band, which agrees reasonably well with the derived COMPTEL flux of 
1.46$\pm$0.21 $\times$ 10$^{-5}$ ph cm$^{-2}$ s$^{-1}$. 
The error on the extrapolated X-ray flux is estimated by using the 1\sig\ 
error on the measured power-law index of \citet{Takahashi09}. The comparison of 
the extrapolation of the X-ray spectrum to the COMPTEL fluxes is graphically 
shown in Fig.~\ref{ls5039_CF_specs}. 

\begin{table}[htb]
\caption[]{
The spectral parameters for the Fermi/LAT sources
as given in the 2nd Fermi Catalog (Nolan et al. 2012). The source names, 
the integral fluxes between 100 and 300~MeV (``F.-flux'' in units of 
10$^{-7} ph\,cm^{-2}\,s^{-1}$), 
the corresponding power-law index (``PL''), 
the spectral shape (``S.-Type''; PL: power-law, LogP: LogParabola, 
PLEC: power-law with exponential cutoff) as well as the spectral extrapolations 
into the COMPTEL 10 to 30 MeV band (``C.-flux'' in units of 
10$^{-6} ph\,cm^{-2}\,s^{-1}$) are listed.
}
\begin{flushleft}
\begin{tabular}{ccccc}
\hline\noalign{\smallskip}
Source             & F.-flux    & PL  & S.-Type  & C.-flux  \\
                   &           &      &          & (10-30)  \\ 
\hline\noalign{\smallskip}
HESS J1825-137     & 1.6569    & 1.690  & PL     & 0.812  \\  
2FGL J1823.1-1338c & 2.0191    & 2.221  & LogP.  & 3.357  \\  
LS 5039            & 3.5651    & 2.406  & LogP.  & 9.080  \\  
LAT PSR J1826-1256 & 2.9473    & 2.191  & PLEC   & 4.575  \\  
2FGL J1827.4-1445c & 1.9279    & 2.052  & PL     & 2.173  \\  
2FGL J1819.3-1523  & 1.0836    & 2.416  & LogP.  & 2.824  \\  
\hline\noalign{\smallskip}
Sum         &               &        &          & 22.82  \\  
\end{tabular}\end{flushleft}
\label{fermi_srcs}
\end{table}

\subsection{\label{variability}Orbit-resolved analyses}

\subsubsection{The {\it INFC} and {\it SUPC} orbital periods}

In recent years it became obvious that the microquasar LS~5039 is a significant \gray\ emitter 
\citep{Aharonian05}, whose \gray\ radiation is modulated along its binary orbit of 
\sm3.9 days \citep{Aharonian06a,Abdo09}. This orbital modulation provides a unique signature
to identify an unidentified \gray\ source as the counterpart of the microquasar. 

\begin{figure*}[th]
  \sidecaption
    \includegraphics[width=12cm]{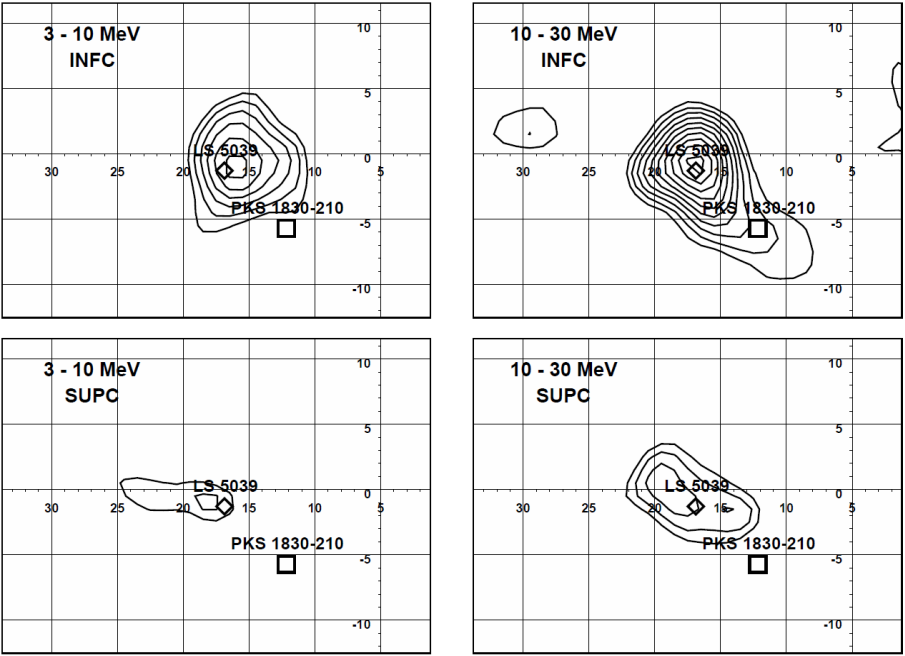}
\caption{
COMPTEL 3-10 and 10-30 MeV significance maps of the LS~5039 sky region, 
generated in galactic coordinates (l, b). The maps give the evidence for GRO~J1823-12  
for the orbital phase intervals {\it INFC} (orbital phase 0.45-0.90)
and {\it SUPC} (orbital phase $\leq$ 0.45 and $>$0.9) of LS~5039. 
The contour lines start at a detection significance of 3\sig\  
(1 d.o.f. for a known source) with a step of 0.5\sig. The locations of the simultaneously 
fitted known COMPTEL sources, LS~5039 (diamonds), PKS~1830-210 (square) are indicated. 
The quasar PKS~1622-297 and a possible source at the Galactic Center, both out of the plot region, 
were also taken into account in the simultaneous fitting.
We clearly find a more significant source for the inferior conjunction period. 
For more details on the maps see text.
}
\label{ls5039_i_s_maps}
\end{figure*}

By using the ephemeris of \citet{Casares05}, determined from radial velocity
measurements of the stellar component in 2002 and 2003, \citet{Aharonian06a} measured a
roughly sinusoidal flux variation along the binary orbit for photons above 1~TeV. 
In particular they found, that the bulk of the TeV flux is emitted along roughly 
half of the orbit at the phase interval $\Phi$ \sm\ 0.45 to 0.9. The VHE flux maximum occurs 
near {\it inferior conjunction} ($\Phi =$ 0.716), while the VHE flux minimum occurs 
at $\Phi$ \sm 0.2, slightly after the orbital {\it superior conjunction} at $\Phi$ = 0.058.
To increase statistics for further analyses, they defined two broad phase intervals: 
{\it INFC} ($0.45 < \Phi \leq\ 0.9$), the orbital part around inferior conjunction, and
{\it SUPC} ($\Phi \leq 0.45$ and $\Phi > 0.9$) the orbital part around superior conjunction.
Their phase-resolved analyses showed for {\it INFC} a bright TeV source (0.2 to 10.0 TeV) 
with a hard power-law spectrum (photon index \sm 1.85) and an exponential cutoff at roughly 9~TeV, 
while for {\it SUPC} a much weaker source being consistent with a soft (photon index \sm 2.53)
and continous power-law shape from 0.2 to 10 TeV.

These definitions of orbital phase intervals, {\it INFC} and {\it SUPC} by 
\citet{Aharonian06a} were susbsequently applied e.g. in the analyses of Fermi/LAT 
\gray\ data at 100~MeV to 10~GeV \citep{Abdo09, Hadasch12}, INTEGRAL 
IBIS/ISGRI hard X-ray data at 13 - 250~keV \citep{Hoffmann09}, and {\it Suzaku} X- and hard 
X-ray data from 1 to 60~keV \citep{Takahashi09}. At all these wavelength bands an orbital 
modulation of the microquasar emission was detected, although not 
always in phase with each other. 

Because COMPTEL MeV detections of GRO~J1823-12 are only possible by averaging a significant 
amount of data, e.g. one CGRO phase covering roughly one year, we cannot perform an independent 
periodicity search. However, we can use the known orbit ephemeris of the microquasar 
and subdivide our data in pieces with respect to orbital phase. As the other instruments, 
we used the ephemeris of \citet{Casares05}, P$_{orb} = 3.90603\pm0.00017$ days with 
T$_{0} = 2451943.09\pm0.10$ (HJD) as periastron passage ($\equiv$ phase $\Phi$ = 0.0),
for our orbit-resolved analyses. In a first step we subdivided the COMPTEL MeV data 
by this ephemeris into the two parts: {\it INFC} and {\it SUPC}. 
This resulted - for a given observational period - in two independent sets of data, which 
are then analysed by the standard COMPTEL analysis procedure (see Sect.~2). 
Fig.~\ref{ls5039_i_s_maps} shows significance maps, for which the sum of all COMPTEL data are 
subdivided in the orbital intervals {\it INFC} and {\it SUPC}. 
The maps are generated for the two more reliable high-energy 
COMPTEL bands. We clearly detect for both energy bands a more significant 
source for the inferior conjunction period, despite the smaller 
exposure accumulated in this orbital part ($\Phi =$ 0.45 - 0.9) compared to 
the {\it SUPC} part. This indicates that at least a significant fraction
of this MeV emission is modulated with the \sm3.9~day orbital period of
the binary system, and - in fact - suggests to be in phase with the TeV emission.
This make LS~5039 very likely the counterpart of this MeV emission, 
so far designated GRO~J1823-12.

\begin{figure}[th]
\centering
   \includegraphics[width=\columnwidth]{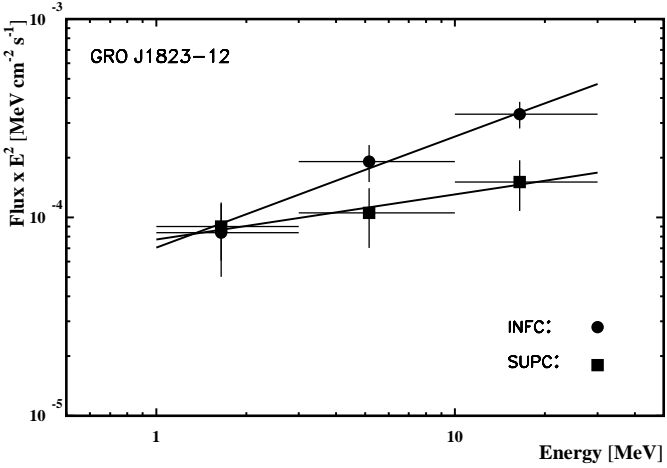}
\caption{
COMPTEL energy spectra, extracted at the location of LS~5039, in an E$^{2} \times$ differential 
flux representation for the sum of all data. Two spectra together with 
their best-fit power-law shapes (solid lines) are shown, representing 
the orbital phase intervals {\it INFC} and {\it SUPC}. The fluxes are 
derived in the 3 standard COMPTEL energy bands (1-3, 3-10, 10-30 MeV). 
The error bars are 1\sig.
}
\label{ls5039_spec_orb}
\end{figure}

As for the orbit-averaged analyses we have extracted fluxes at the location of LS~5039
(Tab.~\ref{ls5039_i_s_flx}) and compiled spectra, which are fitted by power-law 
models (Fig.~\ref{ls5039_spec_orb}).
For both orbital phase regions, {\it INFC} and {\it SUPC} we find hard power-law spectra, 
with the trend of being harder during the {\it INFC} period (Table~\ref{tab_spec_orb}).

A comparison of our COMPTEL 10-30~MeV fluxes to the extrapolated values of the spectra of 
\citet{Takahashi09} for the {\it INFC} as well as the {\it SUPC} period yields comparable values. 
We derive extrapolated fluxes (units $10^{-5}$ ph cm$^{-2}$ s$^{-1}$) 
of 1.5$\pm$0.4 for INFC and 0.57$\pm$0.12 for {\it SUPC}. They are quite comparable to the 
COMPTEL measurements of 2.15$\pm$0.33 for INFC 
and 0.98$\pm$0.28 for {\it SUPC}. The errors on the extrapolated values are estimated by using 
the 1\sig\ errors on the measured X-ray slopes of \citet{Takahashi09}. 

\begin{table}[htb]
\caption{Fluxes, extracted at the location of LS~5039 for the sum of all data, subdivided
into the two orbital phase intervals {\it INFC} ($0.45 < \Phi \leq 0.9$) and 
{\it SUPC} ($\Phi \leq 0.45$ and $\Phi > 0.9$). The flux units are 10$^{-5}$ ph cm$^{-2}$ s$^{-1}$. 
The energy bands are given in MeV. The errors are 1\sig.
} 
\begin{flushleft}
\begin{tabular}{ccccc}
\hline\noalign{\smallskip}
  Period & 1-3 & 3-10 & 10-30   \\
\hline\noalign{\smallskip}
{\it INFC} & 5.41$\pm$2.17 & 4.34$\pm$0.92& 2.15$\pm$0.33 \\
{\it SUPC} & 5.81$\pm$1.90 & 2.39$\pm$0.80& 0.98$\pm$0.28 \\
\hline\noalign{\smallskip}
\end{tabular}\end{flushleft}
\label{ls5039_i_s_flx}
\end{table}

\begin{table}[htb]
\caption[]{Results of the power-law fitting of the COMPTEL spectra  (1-30~MeV)
for the two orbital phase intervals {\it INFC} and {\it SUPC}. 
The errors are 1\sig\ ($\chi^{2}_{min}$ + 2.3 for 2 parameters of interest).
}
\begin{flushleft}
\begin{tabular}{cccc}
\hline\noalign{\smallskip}
Observations & Photon Index & I$_0$ (at 5 MeV) & $\chi^{2}_{min}$   \\
Data & ($\alpha$) & (10$^{-6}$ cm$^{-2}$ s$^{-1}$ MeV$^{-1}$) &   \\ 
\hline\noalign{\smallskip} 
INFC & 1.44$\pm$0.29  & 6.94$\pm$1.61 & 0.23 \\  
SUPC & 1.77$\pm$0.35  & 4.47$\pm$1.31 & 0.06 \\  
\hline\noalign{\smallskip}
\end{tabular}\end{flushleft}
\label{tab_spec_orb}
\end{table}

\subsubsection{Orbital light curve at MeV energies}

To further test on the orbital modulation of the MeV emission, we subdivided 
all COMPTEL data into 5 orbital phase bins. We choose this binning of 0.2 in 
phase because the COMPTEL data cover the time period between July 12, 1991 and 
January 25, 2000. So, the COMPTEL measurements started \sm3500 days before 
and ended \sm375 days before the time T$_{0} = HJD 2451943.09\pm0.10$ 
(equal to February 2, 2001) for which \citet{Casares05} determined 
the applied ephemeris. Applying their period uncertainty ($\Delta$P) of 0.00017~days, 
yields a phase uncertainty $\Delta\Phi$ ($\Delta\Phi$ = ($\Delta$T$\times\Delta$P/P)) 
of \sm0.15 at CGRO VP 5.0 and 0.017 at CGRO VP 907.0, the first and last COMPTEL 
observations of the LS~5039 sky region. On average the phase error is \sm0.08 during 
the COMPTEL mission with respect to the orbital solution of \citet{Casares05}, 
suggesting a phase binning of 0.2. 
Subsequently, we analysed the COMPTEL 10-30~MeV data 
in five orbital phase bins according to the described analysis procedure (see Sect.~2). 
This data selection provides 1) the best source signal in orbit-averaged analyses, 
and 2) the cleanest and most reliable COMPTEL data due to its low 
(compared to the lower COMPTEL energy bands) background, which was 
stable along the COMPTEL mission, e.g. unaffected by satellite reboosts.

\begin{figure}[th]
\centering
   \includegraphics[width=\columnwidth]{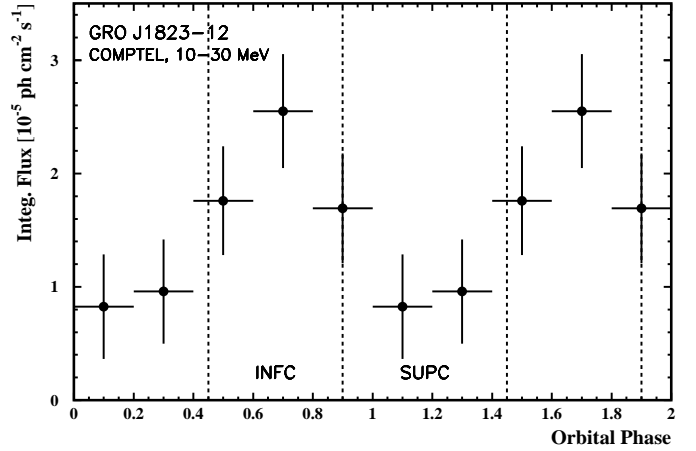}
\caption{
The orbital light curve of LS~5039 in the 10-30~MeV COMPTEL band for the 
sum of all data. The lightcurve is folded with the orbital period of 
\sm3.9 days and given in phase bins of 0.2. The two broader phase periods, 
defined as {\it INFC} and {\it SUPC} are indicated. A flux increase during 
the {\it INFC} period is obvious. In the phase bin containing the inferior conjunction the source 
is roughly three times brighter than in the phase bin containing the 
superior conjunction. In general the 10-30~MeV \gray\ lightcurve 
is consistent in phase and amplitude with the one at TeV \grays.    
}
\label{ls5039_lc_orb}
\end{figure}

We found evidence for an orbital modulation of the COMPTEL 10-30~MeV emission. 
The orbital light curve is shown in Fig.~\ref{ls5039_lc_orb} and the derived 
flux values are given in Table~\ref{ls5039_orb_flx}. A fit of the light curve, 
assuming a constant flux, results in a mean flux value of 
(1.52$\pm$0.21)$\times$$10^{-5} ph\, cm^{-2}\, s^{-1}$ 
with a $\chi^{2}$-value of 8.34 for 4 degrees of freedom. This converts 
to a probability of 0.08 for a constant flux or of 0.92 ($\equiv$1.75\sig) 
for a variable flux. 

Although, we cannot unambigiously prove an orbital 
modulation of the COMPTEL 10-30~MeV emission, the evidence is high.  
In addition to this formal 92\% variability indication, the 
COMPTEL light curve exactly follows the trend in phase and shape 
as is found in other energy bands. In particular
the light curve is in phase with the modulation in X-ray, hard X-ray, and TeV energies, 
i.e. a brighter source near inferior conjunction ($\Phi$ = 0.716) and a weaker one near 
superior conjunction ($\Phi$ = 0.058). The 
measured flux ratio of about a factor of 3 is roughly compatible to the flux ratios 
observed in the X-ray \citep[e.g.][]{Takahashi09} and TeV-bands \citep{Aharonian06a}.
However, the modulation at the MeV band is in anticorrelation to the \gray\ band at 
energies above 100~MeV, observed by Fermi/LAT, where the source is brightest near 
superior conjunction and weakest near inferior conjunction by a flux ratio of 3 to 4
\citep{Abdo09}. By using just the flux values of the phase bins including 
superior (0.0 - 0.2) and inferior (0.6 - 0.8)
conjunction of Table~\ref{ls5039_orb_flx}, we derive a significance of \sm2.5\sig\ 
for a change in flux. This behaviour provides strong evidence for GRO~J1823-12
being, at least for a significant part, the counterpart of the microquasar LS~5039.

\begin{table}[htb]
\caption{Fluxes of a source at the location of LS~5039 for the sum of all COMPTEL data 
in the 10-30~MeV band along the binary orbit in orbital phase bins of 0.2.
The flux unit is 10$^{-5}$ ph cm$^{-2}$ s$^{-1}$.
The errors are 1\sig.
} 
\begin{flushleft}
\begin{tabular}{ccccc}
\hline\noalign{\smallskip}
Orbital Phase  & 10-30~MeV flux   \\
\hline\noalign{\smallskip}
0.0 - 0.2 & 0.83$\pm$0.46 \\
0.2 - 0.4 & 0.96$\pm$0.46 \\
0.4 - 0.6 & 1.76$\pm$0.48 \\
0.6 - 0.8 & 2.55$\pm$0.50 \\
0.8 - 1.0 & 1.69$\pm$0.48 \\
\hline\noalign{\smallskip}
\end{tabular}\end{flushleft}
\label{ls5039_orb_flx}
\end{table}

\begin{figure*}[t]
 \sidecaption
   \includegraphics[width=12cm]{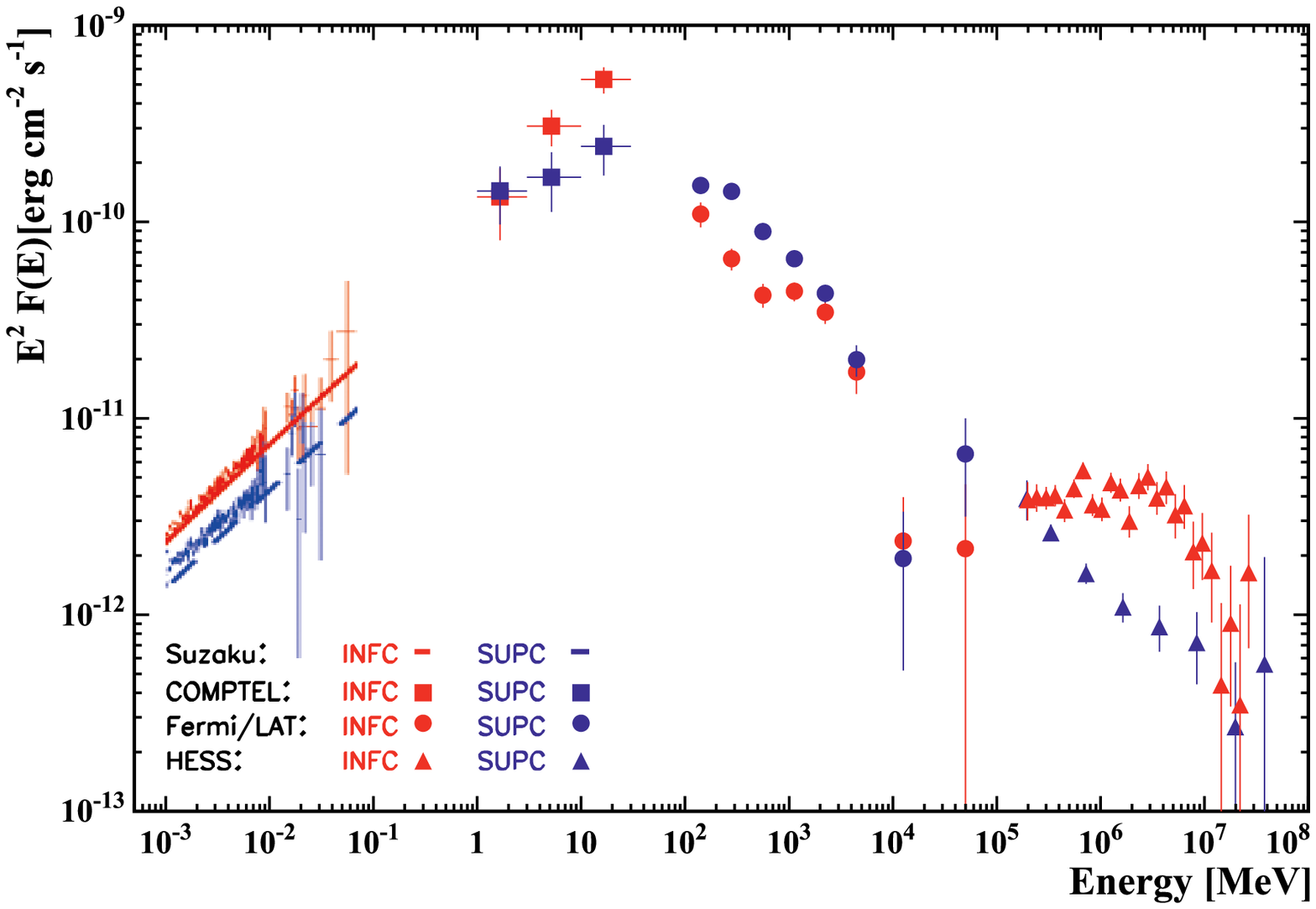}
\caption{
The X-ray to TeV \gray\ SED of LS~5039. The COMPTEL soft \gray\ 
 spectra (sum of all data) for the {\it INFC} 
and {\it SUPC} orbital phases are combined with similar spectra  
from the X-ray \citep[{\it Suzaku},][]{Takahashi09}, 
the MeV/GeV  \citep[Fermi/LAT,][]{Hadasch12}, and 
the TeV band \citep[HESS,][]{Aharonian06a} to build a high-energy SED of LS~5039.
The SED shows that 1) the emission maximum at energies above 1~keV and 2)
a switch in radiation dominance is occuring a MeV energies. 
We like to note that the lines -- solid and dashed -- in the X-ray spectra 
represent the model calculations of \citet{Takahashi09} on the 
emission pattern of LS~5039. They do not present a fit to their measured X-ray spectra. 
}
\label{ls5039_he_sed}
\end{figure*}

\section{ The SED of LS~5039 from X-rays to TeV \grays} 

The peculiar radiation behaviour of LS~5039 is studied across the whole range of the 
electromagnetic spectrum. At high-energies the observational picture is available
in the X- and hard X-ray band (i.e. \sm 1 to 200~keV) and at \grays\ above 100~MeV, 
yielding a significant observational gap at the transition range from the X-rays 
to the \grays. Our analyses of the MeV data of GRO~J1823-12,
in particular the orbit-resolved ones, provide strong evidence for being 
the counterpart of the high-mass X-ray binary LS~5039.   
Fig.~\ref{ls5039_he_sed} shows the high-energy -- X-rays to TeV \grays\ --
SED of the microquasar. We combine our 1-30~MeV spectra, collected for the 
{\it INFC} and {\it SUPC} parts of the orbit, with the similarly collected spectra 
at X-ray, GeV and TeV energies, by assuming that the 
MeV emission is solely due to LS~5039. This may not be completely 
true since other \gray\ sources located within the 
COMPTEL error location region may contribute on a low level.

The COMPTEL measurements fill in a significant part of a yet unknown region of 
the SED, providing new and important information on the emission pattern of LS~5039.
The SED shows that the emission maximum of LS~5039 occurs at MeV energies, 
i.e. between 10 and 100~MeV, and that the dominance in radiation 
between the {\it INFC} and {\it SUPC} orbital periods changes 
between 30 and 100~MeV, i.e. at the transistion region 
between the COMPTEL and Fermi/LAT bands. 
While for {\it SUPC} the SED suggests a kind of smooth transistion from COMPTEL 
to Fermi/LAT, it indicates a kind of complicated transition for the {\it INFC} period. 
For {\it INFC} the SED suggests a strong spectral break between the two bands 
with a drop in flux by a factor of 5 between 30~MeV and 100~MeV. 

In general, the COMPTEL measurements enlighten an interesting region of the LS~5039 
SED where the maximum of the high-energy emission of LS~5039 occurs and where 
significant spectral changes are happening. Therefore these COMPTEL results 
add important information on the emission pattern of LS~5039, 
and so provide additional constraints for the modelling of the microquasar.

\section{\label{discussion}Discussion}

In recent years, after new \gray\ instruments -- in particular 
the ground-based Cherenkov telescopes HESS, MAGIC and VERITAS and the
\gray\ space telescope {\it Fermi} -- became operational, a new
class of \gray\ emitting objects emerged: 
the ``\gray\ binaries''. 
Many binary systems, even of different types
like microquasars, colliding wind binaries, millisecond pulsars in binaries 
and novae, are up to now confirmed emitters of \gray\ radiation 
above 1~MeV (see e.g. \citet{Dubus13} for a recent review). 
Confirmed VHE ($>$100~GeV) emission is yet known 
for five binary systems, among them the microquasar candidate LS~5039. 

LS~5039 is detected at \gray\ energies above 100~MeV by Fermi/LAT 
\citep{Abdo09, Hadasch12} and above 100~GeV by HESS  
\citep{Aharonian05, Aharonian06a} showing a remarkable behaviour
in both bands:  the flux is modulated along its binary orbit, however, in anticorrelation  
for the different waveband regions. Its \gray\ SED is generally ``falling'' from 100~MeV to
TeV-energies. However, showing different shapes depending on 
binary phase. Its general behaviour is not yet understood. Open questions are
for example ``Is LS~5039 a black-hole binary or a neutron-star binary?'', 
which relates to the question, whether the energetic particles necessary 
for the observed high-energy emission are accelerated in a microquasar jet or 
in a shocked region where the stellar and the pulsar wind collide, and ``How this 
translates to the intriguing observed behaviour''. 

Because LS~5039 is now an established \gray\ source at energies above 
100 MeV, for which crucial spectral and timing infos became available in
recent years, we reanalysed all available COMPTEL data of this sky region 
by assuming that the known but unidentified COMPTEL 
source GRO J1823-12 is the counterpart of LS~5039.

As in previous analyses \citep[e.g.][]{Collmar03}, we found a significant 
MeV source being consistent with the sky position of LS~5039, 
which in the COMPTEL 10-30 MeV band shows a kind of stable and steady 
flux level. By extrapolating the measured Fermi/LAT spectra 
of Fermi-detected \gray\ sources in the COMPTEL error box into the COMPTEL band, 
we find LS~5039 to be the most likely counterpart. This identification is supported by
extrapolating the measured {\it Suzaku} X-ray spectra into the MeV band. Assuming 
no spectral breaks from keV to MeV energies, the extrapolations reach flux levels similar to the 
measured COMPTEL fluxes, for the orbit-averaged (see Fig.~\ref{ls5039_CF_specs}) 
as well as orbit resolved analyses. 

Orbit-resolved analyses provide the most convincing evidence for GRO~J1823-12 
being the counterpart of LS~5039. A subdivision of the COMPTEL data into the so-called 
{\it INFC} and {\it SUPC} parts of the binary orbit results in a flux change of the 
MeV source. The flux measurements in the 10-30 and 3-10 MeV band provide evidence at 2.7\sig\ 
and at 1.6\sig\ respectively, that the fluxes for the two orbital phases are different. 
The combined evidence is 3.1\sig.
The analyses of the COMPTEL 10-30~MeV data, subdivided in orbital phase bins 
of 0.2, results in a variable MeV flux. The light curve is consistent in phase, 
shape and amplitude with the ones observed at X-ray and TeV \gray\ energies, 
although the statistical evidence for variability  is only 92\%. 
We like to note that the errors on the fluxes always result from the fitting process 
combining four \gray\ sources and two diffuse galactic emission models. 

By assuming that the fitted fluxes at the location of LS~5039 are solely due to 
LS~5059, i.e. neglecting possible unresolvable contributions of further (Fermi)
\gray\  sources,  we added our MeV flux measurements -- although not simultaneous 
-- to the measured high-energy SED of LS~5039 (Fig.~\ref{ls5039_he_sed}).
The SED shows that the major energy output of LS~5039 occurs at MeV energies, i.e. between 
10 and 100~MeV, for both the {\it INFC} as well as {\it SUPC} orbital parts,
the latter being the less luminous one. 
Both spectra are showing a broad spectral turnover from a harder spectrum below 30 MeV 
to a softer one above 100 MeV, resembling the IC-peaks in COMPTEL-detected 
blazar spectra, e.g. 3C~273 \citep{Collmar00b} or PKS~0528+134 \citep{Collmar97}. 
The dominance of the MeV emission during the {\it INFC} part of the orbit proves 
that the spectral ``flip-back'' from a {\it SUPC}-dominated emission at energies 
above 100 MeV to an INFC-dominated emission occurs between 30 and 100 MeV.

A spectral modelling of the SED is above the scope of this paper. However, we like to 
mention that a changing Compton scattering angle effect may account generally 
for the observed emission pattern. 
If one assumes that the collision of winds in a neutron star binary system
provides an isotropic and relativistic 
plasma and that the high-energy emission is due to inverse-Compton scattering 
of the stellar photons, typically $\sim$9~eV for LS~5039, 
by the relativistic particles, one may need to 
take the Compton-scattering angular cross section into account. 
For the {\it INFC} orbital phase the COMPTON-scatter angles for the photons reaching us 
are on average smaller than for the {\it SUPC} part. For {\it INFC} the primary photons have in 
principle the ``same'' direction than the IC-scattered photons, while for {\it SUPC} the primary 
photons have the ``opposite'' direction than the scattered photons, resulting on average in
larger (SUPC) and smaller (INFC) scatter angles of the IC-process. Because the 
Compton cross section is dependent on the Compton scatter angle, this may result in 
different spectra of the observed radiation. A larger Compton scatter angle results in
a higher energy of the scattered photons but in a lower probability, i.e. lower flux, 
and a smaller Compton scatter angle in a smaller photon energy but with a higher 
probability, i.e. a higher flux. So, the plasma cooling is more effective 
for lower photon energies under small collision angles, resulting in
an {\it INFC}-dominated spectrum at lower photon energies. The higher energy photons, however, 
are likely to be generated by large collison angles, which results in a {\it SUPC}-dominated spectrum 
at higher energies. 
If this is the main effect in LS 5039 for the keV to MeV energy range, it could explain our measurements: 
two almost parallel SEDs for {\it SUPC} and {\it INFC}, where for the lower energies (keV to MeV) the 
INFC flux is higher than the {\it SUPC} one, but for the higher energies (say above 100 MeV)
 the {\it INFC} flux is lower than the {\it SUPC} one.  

At higher energies, say above 10~GeV, other effects, like pair production and adiabatic cooling
\citep[e.g.][]{Takahashi09} will 
take over, the former leading to a higher {\it INFC} flux, while the latter leads to higher plasma energies 
at {\it INFC}, thereby causing the spectral difference between {\it INFC} and {\it SUPC} observed at TeV energies 
by HESS. 

\section{\label{conclusion}Summary and conclusion}

LS~5039 is now an established \gray\ source at energies above 
100 MeV, for which crucial spectral and timing infos became available in
recent years. Because LS~5039 is spatially coincident with the known but unidentfied 
COMPTEL source GRO~J1823-12, we reanalysed the COMPTEL data of this sky region  
by assuming that LS~5039 is the counterpart of GRO J1823-12. We report the data 
analysis work in this paper. 

We provide strong evidence for LS~5039 being the counterpart of GRO~J1823-12. 
Individual statistical tests are not totally convincing by reaching individually up 
to 3\sig\ only. However from the sum of the analyses, we conclude that LS~5039 is 
-- at least for the major fraction -- 
the counterpart of the MeV source. The derived absolute flux values, 
fitted at the sky position of LS~5039 and assumed to be from LS~5039, 
may slightly be too high because we cannot exclude
underlying emission from further \gray\ sources which COMPTEL is unable to resolve. From 
the orbital lightcurve, we estimate them to be up to one third at most by 
assuming that the minimum emission is completely due to other sources. For the 
relative fluxes we show that the source is brighter during {\it INFC} than during {\it SUPC}, 
which shows the dominance in radiation for LS~5039 has to flip back from {\it SUPC} 
at energies above 100~MeV to {\it INFC} between 30 and 100~MeV. This is an important result
which provides a significant constraint for the source modelling. 
Adding the COMPTEL fluxes to the high-energy SED shows that the emission maximum of 
LS~5039 is at MeV energies. While the {\it SUPC} emission almost smoothly turns over from the 
Fermi/LAT band to the COMPTEL band, the {\it INFC} emission seems to show a strong break 
between the COMPTEL and the Fermi/LAT band. This behaviour provides another important 
constraint for modelling the \gray\ binary.

In summary, the COMPTEL data have enlighten a part of the yet ``dark'' region
in the SED of the \gray\ binary LS~5039 thereby providing new insights in its emission 
processes by delivering new and important constraints for the source modelling.

\begin{acknowledgements}
We thank D. Hadasch for providing the Fermi/LAT and HESS spectral points of the 
LS~5039 SED. 
S. Zhang acknowledges support from 973 program 2009CB824800 and the Chinese NSFC 11073021, 
11133002 and XTP project XDA04060604.

\end{acknowledgements}


\begin{thebibliography}{}

\bibitem[Abdo et~al.(2009)]{Abdo09}
Abdo, A. A., Ackermann, M., Ajello, M., et al., 2009, ApJ {\bf 706}, L56

\bibitem[Aharonian et~al.(2005)]{Aharonian05}
Aharonian, F., Akhperjanian, A. G., Aye, K. M., et al., 2005, Science {\bf 309}, 746

\bibitem[Aharonian et~al.(2006a)]{Aharonian06a}
Aharonian, F., Akhperjanian, A. G., Bazer-Bachi, A.R., et al., 2006a, A\&A {\bf 460}, 743

\bibitem[Aharonian et~al.(2006b)]{Aharonian06b}
Aharonian, F., Akhperjanian, A. G., Bazer-Bachi, A.R., et al., 2006b, A\&A {\bf 460}, 365

\bibitem[Bloemen et~al.(1994)]{Bloemen94}
Bloemen H., Hermsen W, Swanenburg B.N., et al., 1994, ApJS {\bf 92}, 419

\bibitem[Casares et~al.(2005)]{Casares05}
Casares, J., Rib\'{o}, M., Ribas, I., et al., 2005, MNRAS {\bf 364}, 899

\bibitem[Collmar et~al.(1997)]{Collmar97}
Collmar, W., Bennett, K., Bloemen, H., et al., 1997, A\&A {\bf 328}, 33-42

\bibitem[Collmar et~al.(2000a)]{Collmar00a}
Collmar, W., Sch\"{o}nfelder, V., Strong, A. W., et al., 2000a, AIP Conf. Proceedings {\bf 510}, 591

\bibitem[Collmar et~al.(2000b)]{Collmar00b}
Collmar, W., Reimer, O., Bennett, K., et al., 2000b, A\&A {\bf 354}, 513-521

\bibitem[Collmar(2003)]{Collmar03}
Collmar, W. 2003, in Proc. of 4th AGILE Science Workshop, Frascati (Rome), on 11-13 June 2003, 177

\bibitem[de Boer et~al.(1992)]{Boer92}
de Boer H., Bennett K., Bloemen H., et al., 1992, in Data Analysis in Astronomy IV, 
eds. V. Di Ges\`{u}, L. Scarsi, R. Buccheri, et al., (New York: plenum Press), p. 241

\bibitem[Dubus(2013)]{Dubus13}
Dubus, G., 2013, Astron. Astrophys. Rev. 21:64

\bibitem[Grondin et~al.(2011)]{Grondin11}
Grondin, M.-H., Funk, S., Lemoine-Goumard, M., et al., 2011, ApJ {\bf 738}, 42

\bibitem[Hadasch et~al.(2012)]{Hadasch12}
Hadasch, D., Torres, D. F., Tanaka, T., et al., 2012, ApJ {\bf 749}, 54

\bibitem[Hartman et~al.(1999)]{Hartman99}
Hartman, R. C., Bertsch, D. L., Bloom, et al., 1999, ApJS {\bf 123}, 79

\bibitem[Hoffmann et~al.(2009)]{Hoffmann09}
Hoffman, A. D., Klochkov, D., Santangelo, A., et al., 2009, A\&A {\bf 494}, L37

\bibitem[Motch et~al.(1997)]{Motch97}
Motch C., Haberl F., Dennerl K., Pakull M., Janot-Pacheco E., 1997, A\&A {\bf 323}, 853

\bibitem[Nolan et al.(2012)]{Nolan12}
Nolan, P.L., Abdo, A.A., Ackermann, M., et al., 2012, ApJS 199:31 

\bibitem[Paredes et~al.(2000)]{Paredes00}
Paredes, J. M., Mart\'{i}, J., Rib\'{o}, M., Massi, M., 2000, Science {\bf 288}, 2340

\bibitem[Sch\"onfelder et~al.(1993)]{Schoenfelder93}
Sch\"{o}nfelder, V., Aarts, H., Bennett, K., et al., 1993, ApJS {\bf 86}, 657 

\bibitem[Sch\"onfelder et~al.(2000)]{Schoenfelder00}
Sch\"{o}nfelder, V., Bennett, K., Blom, J. J., et al., 2000, A\&AS {\bf 143}, 145 

\bibitem[Stephenson \& Sanduleak(1971)]{Stephenson71}
Stephenson C. B., Sanduleak N., 1971, Publ. Warner Swasey Obs., 1, 1

\bibitem[Strong et~al.(2001)]{Strong01}
Strong, A. W., Collmar, W., Bennett, K., et al., 2001,  AIP Conf. Proceedings {\bf 587}, 21-25

\bibitem[Takahashi et~al.(2009)]{Takahashi09}
Takahashi, T., Kishishita, T., Uchiyama, Y., et al., 2009, ApJ {\bf 697}, 592 

\bibitem[Weidenspointner et~al.(1999)]{Weidenspointner99}
Weidenspointner, G., Varendorff, M., Bennett K., et al., 1999, ApL\&C {\bf 39}, 193 

\bibitem[Zhang et~al.(2002)]{Zhang02}
Zhang, S., Collmar, W., Bennett, K., et al., 2002, A\&A {\bf 386}, 843 

\bibitem[Zhang et~al.(2008)]{Zhang08}
Zhang, S., Chen, Y.-P., Collmar, W., et al., 2008, ApJ {\bf 683}, 400 

\end{thebibliography}
\end{document}